\documentstyle[12pt,fleqn]{article}

\font\twlgot =eufm10 scaled \magstep1
\font\egtgot =eufm8
\font\sevgot =eufm7

\font\twlmsb =msbm10 scaled \magstep1
\font\egtmsb =msbm8
\font\sevmsb =msbm7

\newfam\gotfam
\def\pgot{\fam\gotfam\twlgot}
\textfont\gotfam\twlgot
\scriptfont\gotfam\egtgot
\scriptscriptfont\gotfam\sevgot
\def\got{\protect\pgot}

\newfam\msbfam
\textfont\msbfam\twlmsb
\scriptfont\msbfam\egtmsb
\scriptscriptfont\msbfam\sevmsb

\def\pBbb{\relax\ifmmode\expandafter\Bb\else\typeout{You cann't use
Bbb in text mode}\fi}
\def\Bb #1{{\fam\msbfam\relax#1}}

\let\Large=\large

\def\op#1{\mathop{{\it\fam0} #1}\limits}

\newcommand{\nm}[1]{\mid {#1}\mid}

\newcommand{\bite}{\begin{itemize}}
\newcommand{\eite}{\end{itemize}}
\newcommand{\benu}{\begin{enumerate}}
\newcommand{\eenu}{\end{enumerate}}
\newcommand{\bde}{\begin{description}}
\newcommand{\ede}{\end{description}}
\newcommand{\bquo}{\begin{quote}}
\newcommand{\equo}{\end{quote}}
\newcommand{\bquot}{\begin{quotation}}
\newcommand{\equot}{\end{quotation}}

\newcommand{\eqref}[1]{(\ref{#1})}
\newcommand{\beq}{\begin{equation}}
\newcommand{\eeq}{\end{equation}}
\newcommand{\ben}{\begin{eqnarray}}
\newcommand{\een}{\end{eqnarray}}
\newcommand{\be}{\begin{eqnarray*}}
\newcommand{\ee}{\end{eqnarray*}}
\newcommand{\bea}{\begin{eqalph}}
\newcommand{\eea}{\end{eqalph}}

\newcommand{\cG}{{\got g}}

\newcommand{\gT}{{\got T}}

\newcommand{\cL}{{\cal L}}

\newcommand{\cE}{{\cal E}}

\newcommand{\cF}{{\cal F}}

\newcommand{\cS}{{\cal S}}

\newcommand{\bL}{{\bf L}}

\newcommand{\al}{\alpha}
\newcommand{\bt}{\beta}
\newcommand{\dl}{\delta}
\newcommand{\la}{\lambda}

\newcommand{\f}{\phi}

\newcommand{\s}{\psi}
\newcommand{\x}{\xi}
\newcommand{\om}{\omega}

\newcommand{\m}{\mu}
\newcommand{\n}{\nu}

\newcommand{\G}{\Gamma}
\newcommand{\e}{\epsilon}
\newcommand{\ve}{\varepsilon}
\newcommand{\th}{\theta}

\newcommand{\si}{\sigma}

\newcommand{\Y}{Y\to X}
\newcommand{\w}{\wedge}

\newcommand{\wt}{\widetilde}

\newcommand{\dr}{\partial}

\newcommand{\mto}{\mapsto}

\newcommand{\ot}{\otimes}
\newcommand{\ap}{\approx}

\let\ssection=\section
\renewcommand{\section}{\setcounter{equation}{0}\ssection}

\newcounter{eqalph}[section]
\newcounter{equationa}[section]
\newcounter{example}[section]
\newcounter{remark}[section]
\newcounter{theorem}[section]
\newcounter{proposition}[section]
\newcounter{lemma}[section]
\newcounter{corollary}[section]
\newcounter{definition}[section]

\def\theremark{\arabic{section}.\arabic{remark}}

\def\thedefinition{\arabic{section}.\arabic{definition}}

\newenvironment{rem}{\refstepcounter{remark} \medskip\noindent{\bf Remark
\theremark.}\small}{{\Large $\bullet$} \bigskip }

\newenvironment{eqalph}{\stepcounter{equation}
\setcounter{equationa}{\value{equation}}
\setcounter{equation}{0}

\begin{eqnarray}}{\end{eqnarray}
\setcounter{equation}{\value{equationa}}}

\newcommand{\mar}[1]{}

\begin{document}
\hbox{}

{\parindent=0pt

{\large\bf Energy-momentum conservation laws in gauge theory with
broken gauge invariance} 
\bigskip

{\sc  Gennadi
Sardanashvily}
\bigskip

\begin{small}
Department of Theoretical Physics, Physics Faculty, Moscow State
University, 117234 Moscow, Russia

E-mail: sard@grav.phys.msu.su

\bigskip

{\bf Abstract.}

If a Lagrangian of gauge theory of internal symmetries is not
gauge-invariant, the energy-momentum fails to be  
conserved in general.
\end{small}
}

\section{Introduction}

We follow the geometric formulation of
classical field theory where fields are represented by sections of a
fibre bundle $Y\to X$, coordinated by $(x^\la,y^i)$ (see
\cite{book,book00,epr02} for a survey).  Then gauge
transformations are defined as automorphisms of $Y\to X$. 
A gauge transformation is called internal if it is a vertical automorphism
of $Y\to X$, i.e., is projected onto the identity morphism of the
base $X$. 
To study the
invariance conditions and conservation laws, it suffices to consider
one-parameter groups of gauge transformations. Their infinitesimal
generators are projectable vector fields 
\mar{e1}\beq
u=u^\la(x^\m)\dr_\la +u^i(x^\m,y^j)\dr_i \label{e1}
\eeq
on a fibre bundle $Y\to X$. In particular, generators of internal gauge
transformations are vertical vector fields
\mar{e2}\beq
u=u^i(x^\m,y^j)\dr_i. \label{e2}
\eeq

We are concerned with a first order Lagrangian field theory. 
Its configuration space is the
first order jet manifold $J^1Y$ of $Y\to X$, coordinated by
$(x^\la,y^i,y^i_\la)$. A first order Lagrangian
is defined as a density
\mar{cc201}\beq
L=\cL(x^\la,y^i,y^i_\la)\om, \qquad \om=dx^1\w\cdots\w dx^n, \qquad
n=\dim X, \label{cc201}
\eeq
on $J^1Y$. A Lagrangian $L$ is invariant under
a one-parameter group of gauge transformations generated by a vector
field $u$ (\ref{e1}) iff its Lie derivative 
\mar{e100}\beq
\bL_{J^1u}L=J^1u\rfloor dL +d(J^1u\rfloor L) \label{e100}
\eeq
along the jet
prolongation $J^1u$ of $u$
vanishes. In this case, the first variational
formula of the calculus of variations
leads on-shell to the weak conservation law 
\mar{e13}\beq
d_\la\gT^\la_u\ap 0 \label{e13}
\eeq
of the
current
\mar{Q30}\ben
&&\gT_u =\gT_u^\la\om_\la, \qquad \om_\la=\dr_\la\rfloor\om, \nonumber\\
&& \gT_u^\la =(u^\m y^i_\m-u^i)\dr^\la_i\cL -u^\la\cL, 
\label{Q30}
\een
along $u$. In particular, the current $\gT_u$ (\ref{Q30}) along a vertical
vector field $u$ (\ref{e2})  reads
\mar{b374}\beq
\gT^\la_u =-u^i\dr^\la_i\cL. \label{b374}
\eeq
It is called the Noether current.

It is readily observed that 
\mar{e4}\beq
\gT_{u+u'}=\gT_u +\gT_{u'}. \label{e4}
\eeq
Note that any projectable vector field $u$
(\ref{e1}), projected onto 
the vector field $\tau=u^\la\dr_\la$ on $X$, can be written as the sum
\mar{e3}\beq
u=\wt\tau +(u-\wt\tau) \label{e3}
\eeq
of some lift $\wt\tau=u^\la\dr_\la +\wt\tau^i\dr_i$
of $\tau$ onto $Y$ and the vertical vector field 
$u-\wt\tau$ on $Y$. 
The current $\gT_{\wt\tau}$ (\ref{Q30}) along a lift $\wt\tau$ onto $Y$
of a vector field $\tau=\tau^\la\dr_\la$ on $X$ is said to be the
energy-momentum current \cite{fer,book,got,sard97}. It is linear in
components of a vector field $\tau$ and their partial
derivatives. Therefore, one can think of 
$\gT_{\wt\tau}$ as being a linear differential operator on vector
fields on $X$. Then the decompositions (\ref{e4}) and (\ref{e3}) 
show that any current
$\gT_u$ (\ref{Q30}) along a projectable vector field $u$ on a fibre
bundle $Y\to X$ can be represented by a sum of an energy-momentum
current and a Noether one. 
 
Different lifts $\wt\tau$ and $\wt\tau'$ onto $Y$ of a vector field
$\tau$ on $X$ lead to distinct energy-momentum currents $\gT_{\wt\tau}$
and $\gT_{\wt\tau'}$, whose difference $\gT_{\wt\tau}-\gT_{\wt\tau'}$
is the Noether current along the vertical vector field
$\wt\tau-\wt\tau'$ on $Y$. The problem is that, in general, there
is no canonical lift onto $Y$ of vector fields on $X$, and one
can not take the Noether part away from an energy-momentum current.
Therefore, if a Lagrangian is not invariant under vertical gauge
transformations, there is an obstruction for energy-momentum
currents to be conserved \cite{hyper}.

Note that there exists the category of so called natural fibre bundles
$T\to X$ which admit the canonical lift $\wt\tau$ of any vector 
field $\tau$ on $X$ \cite{kol}. This lift is the infinitesimal generator of a
one-parameter group of general covariant transformations of $T$.
For instance, any tensor bundle
\mar{mos6}\beq
T=(\op\ot^mTX)\ot(\op\ot^kT^*X) \label{mos6}
\eeq
over $X$ is of this type. The canonical lift onto $T$ (\ref{mos6})
of a vector field $\tau$
on $X$ is
\mar{l28}\beq
\wt\tau = \tau^\m\dr_\m + [\dr_\nu\tau^{\al_1}\dot
x^{\nu\al_2\cdots\al_m}_{\bt_1\cdots\bt_k} + \ldots
-\dr_{\bt_1}\tau^\nu \dot x^{\al_1\cdots\al_m}_{\nu\bt_2\cdots\bt_k}
-\ldots]\frac{\dr}{\dr \dot
x^{\al_1\cdots\al_m}_{\bt_1\cdots\bt_k}}. \label{l28}
\eeq
For instance, gravitation theory is a gauge field theory on natural
bundles. Its 
Lagrangians are invariant under general covariant transformations. 
The corresponding conserved energy-momentum current on-shell takes the form
\mar{e6}\beq
\gT_{\wt\tau}^\la\ap d_\m U^{\m\la}, \label{e6}
\eeq
where $U^{\m\la}=-U^{\la\m}$ is the generalized Komar superpotential
\cite{bor,giacqg,book,sard97b}. Other energy-momentum currents
differ from $\gT_{\wt\tau}$ (\ref{e6}) in Noether currents, but they
fail to be conserved because almost all gravitation Lagrangians are not
invariant under vertical (non-holonomic) gauge transformations.

Here, we focus on energy-momentum conservation laws in gauge theory of
principal connections on a principal bundle $P\to X$ with a structure
Lie group $G$. These connections are sections of the
fibre bundle 
\mar{e7}\beq
C=J^1P/G\to X, \label{e7}
\eeq
and are identified to gauge potentials \cite{book,book00,epr02}. 
The well-known result claims that, if $L$ is a gauge-invariant 
Lagrangian on $J^1C$ in the presence of a background
metric $g$, we have the familiar covariant conservation law
\mar{C112}\beq
\nabla_\la (t^\la_\m\sqrt{\nm g})\ap 0\label{C112}
\eeq
of the metric energy-momentum tensor
\mar{e40}\beq
t^\m_\bt\sqrt{\nm g}=2 g^{\m\al}\dr_{\al\bt}\cL, \label{e40}
\eeq
where $\nabla$ is the covariant derivative with respect
to the Levi--Civita connection of the background metric
$g$ \cite{got}. Moreover, other energy-momentum conservation laws differ from
(\ref{C112}) in superpotentials terms $d_\m d_\la U^{\m\la}$.
Here, we show that the conservation law (\ref{C112}) locally holds
without fail. However, no energy-momentum current is conserved if a
principal bundle $P$ is not trivial and a 
Lagrangian of gauge theory on $P$ is not gauge-invariant. 

Two examples of
non-invariant Lagrangians are examined. The first one is the
Chern--Simons Lagrangian  whose Euler--Lagrange operator is
gauge-invariant. In this case, we have a conserved quantity, but it
differs from an energy-momentum current. Another 
example is the Yang--Mills Lagrangian in the presence of a
background field, e.g., a Higgs field.

\section{Lagrangian conservation laws}

The first variational formula provides the following universal
procedure for the study of Lagrangian conservation laws in 
classical field theory. 

\begin{rem}
Let $J^2Y$ be the second order jet manifold coordinated by
$(x^\la,y^i,y^i_\la, y^i_{\la\m})$. Recall the following standard
notation: of the contact form $\th^i=dy^i-y^i_\la dx^\la$,
the horizontal projection
\be
h_0(dx^\la)=dx^\la, \qquad h_0(dy^i)=y^i_\la dx^\la \qquad h_0(dy^i_\m)=
y^i_{\la\m}dx^\la,
\ee
the total derivative 
\be
d_\la =\dr_\la + y^i_\la\dr_i +y^i_{\la\m}\dr^\m_i,
\ee
and the horizontal differential
$d_H=dx^\la\w d_\la$ such that $d_H\circ h_0=h_0\circ d$.
\end{rem}

Let $u$ be a projectable vector field on a fibre bundle $\Y$ and
\mar{e22}\beq
J^1u=u +(d_\la u^i-y^i_\m\dr_\la u^\m)\dr^\la_i \label{e22}
\eeq
its jet prolongation onto $J^1Y$. The Lie derivative (\ref{e100})
of a Lagrangian $L$ along $J^1u$ reads
\mar{04}\beq
\bL_{J^1u}L= [\dr_\la u^\la\cL +(u^\la\dr_\la+
u^i\dr_i +(d_\la u^i -y^i_\m\dr_\la u^\m)\dr^\la_i)\cL]\om. \label{04}
\eeq
The first variational formula states
its canonical decomposition over $J^2Y$:
\mar{bC30'}\ben
&& \bL_{J^1u}L= 
 u_V\rfloor \cE_L + d_Hh_0(u\rfloor H_L) \label{bC30'} =\\
&& \qquad   (u^i-y^i_\m u^\m )(\dr_i-d_\la \dr^\la_i)\cL\om - 
d_\la[(u^\m y^i_\m -u^i)\dr^\la_i\cL -u^\la\cL]\om, \nonumber
\een
where $u_V=(u\rfloor\th^i)\dr_i$, 
\mar{305}\beq
\cE_L= (\dr_i\cL- d_\la\dr^\la_i\cL) \th^i\w\om,
\label{305} 
\eeq
is the Euler--Lagrange operator, and
\mar{303}\beq
H_L=L +\dr^\la_i\cL\th^i\w\om_\la=\dr^\la_i\cL dy^i\w\om_\la +(\cL- 
y^i_\la\dr^\la_i\cL)\om
\label{303}
\eeq
is the Poincar\'e--Cartan form. 

The kernel of the Euler-Lagrange operator $\cE_L$ (\ref{305}) is
given by the coordinate relations
\mar{b327}\beq
\dl_i\cL=(\dr_i- d_\la\dr^\la_i)\cL=0, \label{b327} 
\eeq
and defines the Euler--Lagrange equations. 
Their classical solution is a section
$s$ of the fibre bundle $X\to Y$ whose second order jet prolongation $J^2s$
lives in (\ref{b327}). 

\begin{rem}
Note that different Lagrangians $L$ and $L'$ lead to the same
Euler--Lagrange operator if their difference $L_0=L-L'$ is a 
variationally trivial Lagrangian
whose Euler--Lagrange operator vanishes.
Such a Lagrangian takes the form
\mar{mos11}\beq
L_0=h_0(\e) \label{mos11}
\eeq
where $\e$ is a closed $n$-form on $Y$ \cite{jmp01,epr02}.  We have 
locally $\e=d\si$ and
\be
L_0=h_0(d\si)=d_H(h_0(\si)). 
\ee
\end{rem}

On the shell (\ref{b327}),
the first variational formula (\ref{bC30'}) leads to the weak 
identity 
\mar{J4}\beq
 \dr_\la u^\la\cL +[u^\la\dr_\la+
u^i\dr_i +(d_\la u^i -y^i_\m\dr_\la u^\m)\dr^\la_i]\cL \ap 
- d_\la[\c(u^\m y^i_\m -u^i)\dr^\la_i\cL -u^\la\cL]. \label{J4}
\eeq
If the Lie derivative $\bL_{J^1u}L$ (\ref{04})
vanishes, we obtain the 
weak conservation law $0\ap-d_H\gT_u$ (\ref{e13}) of the current
$\gT_u$ (\ref{Q30}). 
It takes the coordinate form
\mar{K4}\beq
0\ap - d_\la[(u^\m y^i_\m-u^i)\dr^\la_i\cL-u^\la\cL]. \label{K4}
\eeq

\begin{rem}
It should be emphasized that, from the first variational formula, the
symmetry current (\ref{Q30}) is defined 
modulo the terms $d_\m(c^{\m\la}_i(y^i_\nu u^\nu - u^i))$,
where $c^{\m\la}_i$ are arbitrary skew-symmetric functions on $Y$. 
Here we leave aside these boundary terms which are
independent of a Lagrangian.
\end{rem}

The weak conservation law (\ref{K4}) leads to the differential
conservation law
\be
\dr_\la(\gT^\la\circ s)=0
\ee
on a solution $s$ of the Euler--Lagrange equations.
This differential conservation law implies the integral law
\mar{b3118}\beq
\op\int_{\dr N} s^*\gT =0, \label{b3118}
\eeq
where $N$ is a compact $n$-dimensional submanifold of $X$ and 
$\dr N$ denotes its boundary.

\begin{rem}\label{superpt} \mar{superpt}
It may happen that a current $\gT$ (\ref{Q30}) takes the
form
\mar{b381}\beq
\gT= W+ d_HU=(W^\la +d_\m U^{\m\la})\om_\la, \label{b381}
\eeq
where the term $W$ vanishes on-shell ($W\ap 0$) and 
\mar{e15}\beq
U=U^{\m\la}\om_{\m\la}, \qquad \om_{\m\la}=\dr_\m\rfloor\om_\la,  \label{e15}
\eeq
is a horizontal $(n-2)$-form on $J^1Y$. Then one says that
$\gT$ reduces to a superpotential $U$ (\ref{e15})
\cite{fat,book,sard97}. 
In this case, the integral
conservation law (\ref{b3118}) becomes tautological. At the same time, the
superpotential form (\ref{b381}) of $\gT$ implies the following integral
relation
\mar{b3119}\beq
\op\int_{N^{n-1}} s^*\gT = \op\int_{\dr N^{n-1}}
s^*U, \label{b3119}
\eeq
where $N^{n-1}$ is a compact oriented $(n-1)$-dimensional submanifold of $X$
with the boundary
$\dr N^{n-1}$. One can think of this relation as being a part of the
Euler--Lagrange equations written in an integral form.
\end{rem} 

\begin{rem}\label{gensym}\mar{gensym}
Let us consider conservation laws in the case of gauge transformations 
preserving the Euler-Lagrange operator $\cE_L$, but not necessarily a
Lagrangian $L$. Let $u$ be a generator of these transformations.
Then we have 
\be
\bL_{J^2u}\cE_L=0,
\ee
where $J^2u$ is the second order jet prolongation of the vector field $u$.
There is  the relation
\be
\bL_{J^2u}\cE_L=\cE_{\bL_{J^1u}L} 
\ee
\cite{giach90,book}, and we obtain $\cE_{\bL_{J^1u}L}=0$.
It follows that the Lie derivative $\bL_{J^1u}L$ is a variationally
trivial Lagrangian. Hence, it takes the form $h_0(\e)$ (\ref{mos11}). 
Then the weak identity (\ref{J4}) comes to the weak equality
\mar{e18}\beq
h_0(\e)\ap-d_H\gT_u. \label{e18}
\eeq
A glance at this expression shows that 
\mar{e20}\beq
h_0(e)=W+d_H\f, \label{e20}
\eeq
where $W\ap 0$.
Then the equality (\ref{e18}) leads to the weak conservation law
\mar{b3141}\beq
0\ap d_H(\f+\gT_u), \label{b3141}
\eeq
but the conserved quantity $\f+\gT_u$ is not globally defined, unless
$\e$ is an exact form. 
For instance, let $Y\to X$ be an affine bundle. In this case,  
$\e=\ve+d\si$
where $\ve$ is an $n$-form on $X$ \cite{jmp01}. Since the weak equality
$\ve\ap 0$
implies the strong one $\ve=0$, we obtain from the 
expression (\ref{e20}) that 
$\ve$ is also an exact form. Thus, the conserved quantity in the
conservation law (\ref{b3141}) is well defined.
\end{rem}

\begin{rem}\label{bakgr} \mar{bakgr}
Background fields do not live in the dynamic shell (\ref{b327}) and,
therefore, break Lagrangian
conservation laws as follows. Let us consider the product
\mar{C41}\beq
Y_{\rm tot}=Y\op\times_X Y'\label{C41}
\eeq
of a fibre bundle $Y$, coordinated by $(x^\la, y^i)$, whose sections are
dynamic fields and a fibre bundle
$Y'$, coordinated by $(x^\la, y^A)$, whose sections are background
fields which take the background values
\mar{e70}\beq
y^B=\f^B(x), \qquad y^B_\la= \dr_\la\f^B(x). \label{e70}
\eeq
A Lagrangian $L$ is defined
on the total configuration space $J^1Y_{\rm tot}$.
Let $u$ be a projectable vector field on $Y_{\rm tot}$ which also
projects onto $Y'$ because gauge 
transformations of background fields do not depend on the dynamic ones. This
vector field takes the coordinate form
\mar{l68}\beq
u=u^\la(x)\dr_\la + u^A(x^\mu,y^B)\dr_A + u^i(x^\mu,y^B, y^j)\dr_i.
\label{l68} 
\eeq
Substitution of (\ref{l68}) in (\ref{bC30'}) leads to
the first variational formula in the presence of background 
fields
\mar{gm555}\ben
&&\dr_\la u^\la\cL +[u^\la\dr_\la+  u^A\dr_A +
u^i\dr_i +(d_\la u^A -y^A_\m\dr_\la
u^\m)\dr^\la_A + \label{gm555}\\
&&\qquad  (d_\la u^i -y^i_\m\dr_\la
u^\m)\dr^\la_i]\cL =
(u^A-y^A_\la u^\la)\dr_A\cL + \dr^\la_A\cL d_\la (u^A-y^A_\mu u^\mu)
+\nonumber\\
&&\qquad (u^i-y^i_\la u^\la)\dl_i\cL
-d_\la[(u^\m y^i_\m -u^i)\dr^\la_i\cL -u^\la\cL]. \nonumber
\een
Then the following identity
\be
&&\dr_\la u^\la\cL +[u^\la\dr_\la+  u^A\dr_A +
u^i\dr_i +(d_\la u^A -y^A_\m\dr_\la
u^\m)\dr^\la_A + \\
&& \qquad  (d_\la u^i -y^i_\m\dr_\la u^\m)\dr^\la_i]\cL\ap 
(u^A-y^A_\la u^\la)\dr_A\cL + \dr^\la_A\cL d_\la (u^A-y^A_\mu u^\mu) - \\
&&\qquad d_\la[(u^\m y^i_\m -u^i)\dr^\la_i\cL-u^\la\cL]
\ee
holds on the shell (\ref{b327}).
A total Lagrangian $L$ usually is constructed to be invariant under gauge
transformations of the product (\ref{C41}). In this case, we obtain  the 
weak identity 
\mar{l70}\beq
(u^A-y^A_\m u^\m)\dr_A\cL + \dr^\la_A\cL d_\la (u^A-y^A_\mu
u^\mu) \ap d_\la[(u^\m y^i_\m-u^i)\dr^\la_i\cL -u^\la\cL]. \label{l70}
\eeq
in the presence of background  
fields on the shell (\ref{e70}). Given a background field $\f$
(\ref{e70}), there always exists a vector field on a fibre bundle $Y\to
X$ such that the left-hand
side of the equality (\ref{l70}) vanishes. This is the horizontal lift
\be
\wt\tau=\tau^\la(\dr_\la +\G^A_\al\dr_A)
\ee
onto $Y'$ of a vector field $\tau$ on $X$ by means of a connection
$\G$ on $Y'\to X$ whose integral section is $\f$, i.e.,
$\dr_\la\f^A=\G^A_\la\circ\f$. However, the Lie derivative of a
Lagrangian $L$ along this vector field need not vanish.
\end{rem}

\section{Noether conservation laws in gauge theory}

Let $\pi_P :P\to X$ be
a principal bundle  with a structure
Lie group $G$ which acts on $P$ on the right 
\mar{1}\beq
R_g : p\mto pg, \quad p\in P,\quad g\in G. \label{1}
\eeq
A principal bundle $P$ is equipped with a bundle atlas
$\Psi_P=\{(U_\al,\psi^P_\al\}$
whose trivialization morphisms $\s_\al^P$
obey the equivariance condition
\mar{1119}\beq
\s_\al^P(pg)=\s_\al^P(p)g, \qquad \forall g\in G,
\qquad \forall p\in\pi^{-1}_P(U_\al). \label{1119}
\eeq

A gauge transformation in gauge theory on a principal bundle $P\to X$ is
defined as an automorphism $\Phi_P$ of $P\to X$
which is equivariant under the canonical
action (\ref{1}), i.e.,
$R_g\circ\Phi_P=\Phi_P\circ R_g$ for all $g\in G$.
The infinitesimal generator of a one-parameter group of
these gauge transformations is a $G$-invariant vector field $\x$ on $P$. It
is naturally identified to a section 
of the quotient $T_GP=TP/G$ of the tangent bundle $TP\to P$
by the canonical action $R_G$ (\ref{1}).
Due to the equivariance condition (\ref{1119}), any bundle atlas $\Psi_P$
of $P$ yields the
associated bundle atlase $\{U_\al,T\psi^P_\al/G)\}$ of $T_GP$. 
Given a basis $\{e_p\}$ for the
right Lie algebra $\cG_r$ of the group $G$, let $\{\dr_\la, e_p\}$ 
be the corresponding
local fibre bases for the vector bundles $T_GP$. Then a section $\x$
of $T_GP\to X$ reads
\mar{e10}\beq
\x =\x^\la\dr_\la + \x^p e_p. \label{e10}
\eeq
The infinitesimal generator
of a one-parameter group  of vertical gauge transformations is a
$G$-invariant vertical vector field on $P$ identified
to a section $\x=\x^p e_p$ of the quotient 
\mar{b1.205}\beq
V_GP=VP/G\subset T_GP \label{b1.205}
\eeq
of the vertical tangent
bundle $VP$ of $P$ by the canonical action $R_G$ (\ref{1}).

The Lie bracket of two sections
$\x$ and $\eta$ 
of the vector bundle $T_GP\to X$ reads
\mar{1129}\beq
  [\x,\eta ]=(\x^\m\dr_\m\eta^\la - \eta^\m\dr_\m\x^\la)\dr_\la
  +(\x^\la\dr_\la\eta^r - \eta^\la\dr_\la\x^r +
c_{pq}^r\x^p\eta^q) e_r, \label{1129}
\eeq
where $c_{pq}^r$ are the structure constants of the 
Lie algebra ${\got g}_r$. Putting $\xi^\la=0$ and $\eta^\m=0$, we
obtain the Lie bracket 
\mar{1129'}\beq
[\x,\eta]= c_{pq}^r\x^p\eta^q e_r \label{1129'}
\eeq
of sections of the vector bundle $V_GP\to X$.
A glance at the expression (\ref{1129'}) shows
that the typical fibre of $V_GP\to X$ is the 
Lie algebra ${\got g}_r$. The structure group $G$ acts on ${\got g}_r$ by the 
adjoint representation. 

A principal connection on a principal
bundle $P\to X$ is defined as a global section $A$ of the affine
jet bundle $J^1P\to P$ which is
equivariant under the right action (\ref{1}), i.e.,
\mar{ee}\beq
J^1R_g\circ A= A\circ R_g, \qquad \forall g\in G. \label{ee}
\eeq
Due to this equivariance condition, there is
one-to-one correspondence between the principal connections
on a principal bundle $P\to X$ and the global sections $A$ of the
quotient $C$ (\ref{e7})
of the first order jet manifold $J^1P$ of a principal bundle $P\to
X$ by the jet prolongation of the canonical action $R_G$ (\ref{1}). 
The quotient $C$ (\ref{e7}) is an affine bundle over $X$. Given a
bundle atlas $\Psi_P$ of $P$, it is
provided with bundle coordinates
$(x^\la,a^q_\la)$ such that $A^q_\la=a^q_\la\circ A$ are coefficients
of the familiar local connection form  
$A^q_\la dx^\la\ot e_q$ on $X$, i.e., $a^q_\la$ are coordinates of
gauge potentials. Therefore $C$ (\ref{e7}) is called the connection
bundle. Gauge transformations of $P$ generated by the vector field
(\ref{e10}) induce gauge transformations of $C$ whose 
generator is 
\mar{e11}\beq
\x_C=\x^\la\dr_\la+ (\dr_\m\x^r +
c^r_{pq}a^p_\m\x^q-a^r_\la\dr_\m\x^\la)\dr^\m_r. \label{e11}
\eeq

The configuration space of gauge theory is the first order jet manifold
$J^1C$ of $C$ coordinated by $(x^\la,a^q_\la,a^q_{\la\m})$. It admits
the canonical splitting over $C$ which takes the coordinate form
\mar{296'}\beq
a_{\la\m}^r = \frac12(\cF^r_{\la\m} +\cS^r_{\la\m})=
\frac{1}{2}(a_{\la\m}^r + a_{\m\la}^r
  - c_{pq}^r a_\la^p a_\m^q) + \frac{1}{2}
(a_{\la\m}^r - a_{\m\la}^r +
c_{pq}^r a_\la^p a_\m^q). \label{296'}
\eeq

Let $L$ be a 
Lagrangian on $J^1C$. One usually requires of $L$ to
be invariant under vertical gauge transformations. It means 
that the Lie derivative ${\bf L}_{J^1\x_{CY}} L$ of $L$
along the jet prolongation (\ref{e22}) of any vertical vector field 
\mar{C76'}\beq
\x_C=
(\dr_\la \xi^r+c^r_{qp}a^q_\la\xi^p)\dr^\la_r
\label{C76'}
\eeq
on $C$ vanishes. Coefficients $\x^q$ of this
vector field play the role of gauge parameters. Then we come to the
well-known Noether conservation law. The key point is that, since the
vector fields (\ref{C76'}) depends on derivatives of gauge parameters, 
the Noether
current in gauge theory reduces to a superpotential as follows.

The first variational formula
(\ref{bC30'}) leads to the strong equality
\mar{b3109}\beq
0=(\dr_\m \xi^r+c^r_{qp}a^q_\m\xi^p)\dl_r^\m\cL +
d_\la[(\dr_\m \xi^r+c^r_{qp}a^q_\m\xi^p)\dr^{\la\m}_r\cL]. \label{b3109}
\eeq
Due to the arbitrariness of gauge parameters $\x^p$, this equality falls
into the system of equalities
\mar{2110}\bea
&& c_{pq}^r(a_\m^p\dr_r^\m\cL + a^p_{\la\m} \dr_r^{\la\m}\cL)  =  0,
\label{2110a} \\
&& \dr_q^\m \cL + c_{pq}^r a_\la^p \dr_r^{\m\la}\cL  = 0, \label{2110b} \\
&& \dr_p^{\m\la}\cL + \dr_p^{\la\m}\cL = 0. \label{2110c}
\eea
One can think of them as being the equations for
a gauge-invariant
Lagrangian. As is well known, 
there is a unique solution of these equations in the class of quadratic
Lagrangians. It is 
the conventional Yang-Mills Lagrangian $L_{\rm YM}$ of gauge potentials on
the configuration space $J^1C$. In the presence of a background 
metric $g$ on the base
$X$, it reads
\mar{5.1}\beq
L_{\rm YM}=\frac{1}{4\ve^2}a^G_{pq}g^{\la\m}g^{\bt\n}\cF^p_{\la
\bt}\cF^q_{\m\n}\sqrt{\nm g}\om, \qquad  g=\det(g_{\m\n}), \label{5.1}
\eeq
where $\cF^r_{\la\m}$ are components of the canonical splitting
(\ref{296'}) and $a^G$ is a $G$-invariant bilinear form
on the Lie algebra ${\got g}_r$.

On-shell, the strong equality (\ref{b3109}) becomes the
weak Noether conservation law
\mar{C300}\beq
0\ap d_\la[(\dr_\m \xi^r+c^r_{qp}a^q_\m\xi^p)\dr^{\la\m}_r\cL] \label{C300}
\eeq
of the Noether current
\mar{b3110}\beq
\gT^\la_\x=-(\dr_\m \xi^r+c^r_{qp}a^q_\m\xi^p)\dr^{\la\m}_r\cL. \label{b3110}
\eeq
In accordance with the
strong equalities (\ref{2110b}) and
(\ref{2110c}), the Noether current (\ref{b3110}) is brought
into the superpotential
form
\be
\gT^\la_\x =\x^r\dl_r^\la\cL + d_\m U^{\m\la}, \qquad
U^{\m\la}=\x^p\dr^{\la\m}_p\cL.
\ee
The corresponding integral
relation (\ref{b3119}) reads
\mar{b3223}\beq
\op\int_{N^{n-1}} s^*\gT^\la\om_\la = \op\int_{\dr N^{n-1}}
s^*( \x^p \dr^{\m\la}_p)\om_{\m\la}, \label{b3223}
\eeq
where $N^{n-1}$ is a compact oriented $(n-1)$-dimensional submanifold of $X$
with the boundary
$\dr N^{n-1}$. One can think of (\ref{b3223}) as being the integral relation
between the Noether current (\ref{b3110}) and the gauge field, generated by
this current. In electromagnetic theory seen as a $U(1)$ gauge theory,
the similar relation between an electric current and the electromagnetic
field generated by this current is well known. However, it is free from gauge
parameters due to the peculiarity of Abelian gauge models.

\section{Energy-momentum conservation laws in gauge theory}

Let us turn now to energy-momentum conservation laws in gauge
theory. 

Let $B$ be a principal connection on a
principal bundle $P\to X$. Given a vector field $\tau$ on $X$, there
exists its lift
\mar{e25}\beq
\wt\tau_B=\tau^\la\dr_\la +[\dr_\m(\tau^\la B^r_\la)+c^r_{qp}
a^q_\m (\tau^\la B^p_\la) - a^r_\la\dr_\m \tau^\la]\dr^\m_r. \label{e25}
\eeq
onto the connection bundle $C\to X$ (\ref{e7})
\cite{giach90,book,book00,sard97}. 
Comparing the expressions (\ref{e11}) and (\ref{e25}), one easily
observes that the lift $\wt\tau_B$ is a generator of gauge
transformations of $C$ with gauge parameters $\x^\la=\tau^\la$,
$\x^r=\tau^\la B^r_\la$.

Let us discover the energy-momentum current along the lift (\ref{e25}).
We assume that a Lagrangian $L$ of gauge theory 
also depends 
on a background metric on $X$. This metric is described by a section of
the tensor 
bundle $\op\vee^2 TX$ provided with the holonomic coordinates
$(x^\la,\si^{\m\n})$. Following Remark \ref{bakgr}, we define 
$L$ on the total configuration space 
\mar{e30}\beq
J^1Y=J^1(C\op\times_X \op\vee^2 TX). \label{e30}
\eeq

Given a vector field $\tau$ on $X$, there exists its canonical lift 
\mar{e26}\beq
\wt\tau =\tau^\la\dr_\la + 
(\dr_\n\tau^\al\si^{\n\bt}
+\dr_\n\tau^\bt\si^{\n\al})\dr_{\al\bt} \label{e26}
\eeq
(\ref{l28}) onto the tensor bundle $\op\vee^2TX$. It
is a generator of general
covariant transformations of $\op\vee^2TX$. Combining
(\ref{e25}) and (\ref{e26}), we obtain the lift 
\mar{mos54}\ben
&&\wt\tau_Y =\wt\tau_1+\wt\tau_2= [\tau^\la\dr_\la + (\dr_\n\tau^\al\si^{\n\bt}
+\dr_\n\tau^\bt\si^{\n\al})\dr_{\al\bt}
- a^r_\la\dr_\m \tau^\la\dr^\m_r] + \label{mos54}\\
&& \qquad [\dr_\m(\tau^\la B^r_\la)+c^r_{qp}
a^q_\m (\tau^\la B^p_\la)]\dr^\m_r \nonumber
\een
of a vector field $\tau$ on $X$ onto the
product $Y$.
Note that the decomposition (\ref{mos54}) of the lift $\wt\tau_Y$ is
local. One can think of the
summands $\wt\tau_1$ and $\wt\tau_2$ as being local generators of
general covariant transformations (cf. (\ref{l28})) and vertical gauge
transformations (cf. (\ref{C76'}))
of the product $Y$, respectively.

Let a Lagrangian $L$ on the total configuration space (\ref{e30}) be
invariant under general covariant transformations and vertical gauge
transformations. Hence, 
its Lie derivative along
the vector field $\wt\tau_Y$ (\ref{mos54}) equals zero.
Then using the formula (\ref{l70}) on the shell
\be
\si^{\m\nu}=g^{\m\nu}(x),\qquad \dl^\m_r\cL=0,
\ee
we obtain the weak identity  
\mar{mos52}\beq
0\ap (\dr_\n\tau^\al g^{\n\bt}
+\dr_\n\tau^\bt g^{\n\al} -\dr_\la
g^{\al\bt}\tau^\la)\dr_{\al\bt}\cL - d_\la\gT^\la_B, \label{mos52}
\eeq
where
\mar{b3147}\beq
\gT^\la_B = [\dr^{\la\nu}_r\cL (\tau^\m a^r_{\m\nu}+\dr_\nu\tau^\m a^r_\m)
-\tau^\la\cL] + 
 [-\dr^{\la\nu}_r\cL (\dr_\nu(\tau^\m B^r_\m) + 
c^r_{qp}a^q_\n (\tau^\m B^p_\m)] \label{b3147}
\eeq
is the energy-momentum current along the vector field (\ref{e25}).
The weak identity (\ref{mos52}) takes the form
\mar{b3137}\beq
 0\ap \dr_\la\tau^\m t^\la_\m\sqrt{\nm g} -\tau^\m\{_\m{}^\bt{}_\la\}
t^\la_\bt \sqrt{\nm g} - d_\la\gT^\la_B, \label{b3137}
\eeq
where $t^\la_\m$ is the metric energy-momentum tensor (\ref{e40}) and
$\{_\m{}^\bt{}_\la\}$ are the Christoffel symbols  
of $g$. Accordingly, the current $\gT^\la_B$ (\ref{b3147}) is brought
into the form 
\mar{e41}\beq
 \gT^\la_B=\tau^\m t^\la_\m\sqrt{\nm g} +\tau^\al (B^r_\al-a^r_\al)\dl^\la_r\cL
+d_\m(\tau^\al(B^r_\al- a^r_\al)\dr^{\la\m}_r\cL).
\label{e41}
\eeq
Substituting $\gT^\la_B$ (\ref{e41}) into the weak identity
(\ref{b3137}), we obtain the covariant conservation law (\ref{C112})
independent of the choice of the connection $B$ in the lift (\ref{e25}).

\section{The case of broken gauge invariance}

On a local coordinate chart, the conservation law (\ref{C112}) issues directly
from the local decomposition (\ref{mos54}). Namely, the current $\gT_B$
is decomposed locally into the sum $\gT_{\wt\tau_1} + \gT_{\wt\tau_2}$
of the energy-momentum current $\gT_{\wt\tau_1}$ along the the
projectible vector field $\wt\tau_1$ and the Noether current $\gT_{\wt\tau_2}$
along the vertical vector field $\wt\tau_2$. Since the Noether 
current $\gT_{\wt\tau_2}$ is reduced to a superpotential, it does not
contribute to the energy-momentum conservation law (\ref{C112}) if a
Lagrangian $L$ is invariant under vertical gauge transformations.
However, if $L$ is not gauge-invariant, the conservation law
(\ref{C112}) takes the local form
\mar{e50}\beq
\bL_{J^1\wt\tau_2}L= \tau^\m\nabla_\la (t^\la_\m\sqrt{\nm g}) \label{e50}
\eeq
on each coordinate chart. Of course, one can choose $B=0$ and 
restart the conservation law (\ref{C112}) on a given
coordinate chart without fail.
However, if $P$ is a non-trivial principal bundle, no principal
connection on $P$ vanishes everywhere. In this case, no energy-momentum
of gauge fields is conserved.

Turn now to the above mentioned example of the
Chern--Simons Lagrangian \cite{bor98,book}.

Let $P\to X^3$ be a principal bundle over a 
3-dimensional manifold $X$ whose structure group $G$ is a semisimple Lie
group. The Chern--Simons Lagrangian on the configuration space $J^1C$
of principal connections on $P$ reads
\mar{C113}\beq
L_{\rm CS} =\frac{1}{2k}a^G_{mn}\ve^{\al\la\m} a^m_\al (\cF^n_{\la\m}
-\frac13c^n_{pq}a^p_\la a^q_\m)\om, \label{C113}
\eeq
where $\ve^{\al\la\m}$ is the skew-symmetric Levi--Civita tensor and $k$ is
a coupling constant. In comparison with the Yang--Mills Lagrangian
(\ref{5.1}), the Chern--Simons Lagrangian (\ref{C113}) is independent
of any metric on $X$ and is not 
gauge-invariant. At the same time, the Euler--Lagrange
operator 
\mar{b3140}\beq
\cE_{L_{\rm CS}} =\frac1ka^G_{mn}\ve^{\al\la\m}\cF^n_{\la\m} \th^m_\al\w\om.
\label{b3140}
\eeq
is gauge-invariant. Therefore, let us follow Remark
\ref{gensym} in order to study Lagrangian conservation laws in the
Chern--Simons model.

Given a generator $\x_C$ (\ref{C76'}) of vertical gauge
transformations, we obtain
\mar{b3142}\beq
\bL_{J^1\x_C}L_{\rm
CS}=\frac1ka^G_{mn}\ve^{\al\la\m}\dr_\al\x^m a^n_{\la\m}\om. \label{b3142}
\eeq
Since $C\to X$ is an affine bundle, the Lie derivative
(\ref{b3142}) is brought into the form 
\be
\bL_{J^1\x_C}L_{\rm CS} =d_H\f, 
\ee
where
\be
\f=\frac1ka^G_{mn}\ve^{\al\la\m}\dr_\al\x^m a^n_\m\om_\la 
\ee
is a horizontal 2-form on $C\to X$. Then we obtain the weak
conservation law (\ref{b3141}) where
\be
\gT^\la=-\frac1ka^G_{mn}\ve^{\al\la\m}\xi_C{}^n_\m a^m_\al
\ee
is the N\"other current. Moreover, this conservation law takes the
superpotential form 
\be
0\ap  d_\la(\x^\la\cL_{\rm CS} +d_\m U^{\m\la}),\qquad
U^{\m\la} =\frac2ka^G_{mn}\ve^{\al\m\la}\x^na^m_\al.
\ee

Turn now to the energy-momentum conservation law in the Chern--Simons model. 
Let $\tau$ be a vector field on the base $X$ and $\wt\tau_B$
(\ref{e25}) its lift onto the connection bundle $C$ by means of a
principal connection $B$. We obtain
\be
\bL_{J^1\wt\tau_B}L_{\rm
CS}=\frac1ka^G_{mn}\ve^{\al\la\m}\dr_\al(\tau^\nu B^m_\nu)
a^n_{\la\m}\om.
\ee
Then the corresponding conservation law (\ref{b3141}) takes the form
\be
0\ap -d_\la[\gT^\la_B + \frac1ka^G_{mn}\ve^{\al\la\m}\dr_\al(\tau^\nu
B^m_\nu) a^n_\m], 
\ee
where $\gT_B$ is the energy-momentum current (\ref{b3147}) along the
vector field $\wt\tau_B$.
It follows that the energy-momentum current of the Chern--Simons model
is not conserved 
because the Lagrangian (\ref{C113}) is not gauge-invariant, but 
there exists another conserved quantity.

Another example of non-invariant Lagrangians is a Lagrangian of gauge
fields in the presence of a background field (see Remark \ref{bakgr}).
Let us focus on the physically relevant case of gauge theory with
spontaneous symmetry breaking. It is gauge theory on 
a principal bundle $P\to X$ whose structure group $G$ is reduced to
its closed subgroup $H$, i.e., there exists a principal subbundle
$P^\si$ of $P$ with the structure group $H$. 
Moreover, by the well-known theorem
\cite{book,kob}, there is one-to-one correspondence between
the $H$-principal subbundles $P^\si$ of $P$ and the global
sections $\si$ of the quotient bundle $P/H\to X$. These sections are called
Higgs fields. The total Lagrangian $L$ of gauge potentials and Higgs fields
on the configuration space $J^1(C\times P/H)$ is gauge invariant.
Therefore, we can appeal to Remark \ref{bakgr} in order to obtain the
energy-momentum conservation law of gauge potentials in the presence of
a background Higgs field $\si$. The key point is that, due to the
equivariance condition (\ref{ee}),  any principal 
connection on the reduced bundle $P^\si\to X$ gives rise to a principal
connection $A_\si$ on $P\to X$ whose integral section is the
Higgs field $\si$.
Let us consider the lift 
\mar{eee}\beq
\wt\tau=\tau +u_1 +u_2 \label{eee}
\eeq
onto $C\times P/H$ of a vector
field $\tau$ on $X$ such that: $\tau+u_1$ is the lift
$\wt\tau_{A_\si}$ (\ref{e25}) of $\tau$ onto $C$, and $\tau+u_2$ is
the horizontal lift of $\tau$ onto $P/H$ by means of the connection $A_\si$.
Since the Lagrangian $L$ is gauge-invariant, its Lie derivative along
the vector field $\wt\tau$ (\ref{eee}) vanishes. Therefore, we come to
the weak identity (\ref{l70}) whose left-hand side also vanishes, and
we obtain again the energy-momentum conservation law (\ref{C112}).


\begin{thebibliography}{ederf}

\bibitem{bor} A.Borowiec, M.Ferraris, M.Francaviglia and
I.Volovich,Universality of Einstein equations for the Ricci squared
Lagrangians, {\it Class. Quant. Grav.} {\bf 15} (1998) 43.

\bibitem{bor98} A.Borowiec, M.Ferraris and M.Francaviglia, Lagrangian
symmetries of Chern--Simons theories, {\it J. Phys. A} {\bf 31} (1998) 8823.


\bibitem{fat} L.Fatibene, M.Ferraris and M.Francaviglia, N\"other formalism
for conserved quantities in classical gauge field theories, {\it J.Math. 
Phys.} {\bf 35} (1994) 1644.

\bibitem{fer} M.Ferraris and M.Francaviglia, Energy-momentum tensors and
stress tensors in geometric field theories, {\it J. Math.Phys.} 
{\bf 26} (1985) 1243.

\bibitem{giach90} G.Giachetta and L.Mangiarotti, Gauge-invariant and covariant
operators in gauge theories, {\it Int. J. Theor.Phys.} {\bf 29} (1990) 789.

\bibitem{giacqg} G.Giachetta and G.Sardanashvily,
Stress-energy-momentum of affine-metric gravity. Generalized Komar
superportential, {\it Class. Quant. Grav.} {\bf 13} (1996) L67; E-print
arXiv: gr-qc/9511008, gr-qc/9511040.

\bibitem{book} G.Giachetta, L.Mangiarotti and G.Sardanashvily, {\it New
Lagrangian and Hamiltonian Methods in Field Theory} (World Scientific,
Singapore, 1997).


\bibitem{jmp01} G.Giachetta, L.Mangiarotti and G.Sardanashvily, Cohomology
of the infinite-order jet space and the inverse problem, {\it J. Math. Phys.}
{\bf 42} (2001) 4272.

\bibitem{got} M.Gotay and J.Marsden, Stress-energy-momentum tensors and
the Belinfante--Rosenfeld formula, {\it Contemp. Mathem.} {\bf 132}
(1992) 367.

\bibitem{kob} S.Kobayashi and K.Nomizu, {\it Foundations of Differential
Geometry, Vol.1}  (Interscience Publ., N.Y., 1963).

\bibitem{kol} I.Kol\'a\v{r}, P.Michor and J.Slov\'ak, {\it Natural Operations 
in Differential Geometry} (Springer-Verlag, Berlin, 1993).

\bibitem{book00} L.Mangiarotti and G.Sardanashvily, {\it Connections in
Classical and Quantum Field Theory} (World Scientific, Singapore, 2000).

\bibitem{sard97} G.Sardanashvily, Stress-energy-momentum tensors in 
constraint field theories, {\it J. Math. Phys.} {\bf 38} (1997) 847. 

\bibitem{sard97b} G.Sardanashvily, Stress-energy-momentum conservation law in 
gauge gravitation theory, {\it Class. Quant. Grav.} {\bf 14}  (1997) 1371.

\bibitem{hyper} G.Sardanashvily, Energy conservation laws and
anti-matter, {\it Hyperfine Interactions} {\bf 109} (1997) 117.

\bibitem{epr02} G.Sardanashvily, Ten lectures on jet manifolds in
classical and quantum field theory, E-print arXiv: math-ph/0203040.


\end{thebibliography}
\end{document}